\documentclass[12pt, preprint]{aastex}
\newcommand{\hho}{H$_2$O$\,$}

\newcommand{\kms}{km s$^{-1}\,$}
\newcommand{\vlsr}{V$_{\rm LSR}$}

\newcommand{\lsun}{\mbox{L$_{\sun}$}}
\newcommand{\bm}{beam$^{-1}$}

\shorttitle{\hho megamasers in NGC\,5793}
\shortauthors{Hagiwara et al.}

\begin{document}


\title{VLBI STUDY OF WATER MASER EMISSION IN THE SEYFERT 2 GALAXY NGC\,5793. I: IMAGING BLUESHIFTED EMISSION AND THE PARSEC-SCALE JET}


\author{\sc Yoshiaki Hagiwara}
\affil{Max-Planck-Institut f{\"u}r Radioastronomie, Auf dem H{\"u}gel 69, D-53121, Bonn, Germany}
\email{hagi@mpifr-bonn.mpg.de}
\author{\sc Philip J. Diamond}
\affil{Jodrell Bank Observatory, University of Manchester, Macclesfield Cheshire SK 11 9DL, UK}

\and

\author{\sc Naomasa Nakai and Ryohei Kawabe}
\affil{Nobeyama Radio Observatory, Minamimaki, Minamisaku, Nagano, 384-1305, Japan}


%
\begin{abstract}
We present the first result of VLBI observations of the blueshifted
water maser emission from the type 2 Seyfert galaxy NGC\,5793, which
we combine with new and previous VLBI observations of continuum
emission at 1.7, 5.0, 8.4, 15, and 22 GHz. Maser emission was detected
earlier in single-dish observations and found to have both red- and
blueshifted features relative to the systemic velocity. We could image
only the blueshifted emission, which is located 3.6 pc southwest of
the 22 GHz continuum peak. The blueshifted emission was found to
originate in two clusters that are separated by 0.7 milliarcsecond
(0.16 pc). No compact continuum emission was found within 3.6~pc of the
maser spot. A compact continuum source showing a marginally inverted
spectrum between 1.7 and 5.0 GHz was found 4.2 pc southwest of the
maser position. The spectral turnover might be due to synchrotron
self-absorption caused by a shock in the jet owing to collision with
dense gas, or it might be due to free-free absorption in an ionized
screen possibly the inner part of a disk, foreground to the jet.

The water maser may be part of a maser disk. If so, it would be
rotating in the opposite sense to the highly inclined galactic disk
observed in CO emission. We estimate a binding mass within 1 pc of the
presumed nucleus to be on the order of 10$^7$ $M_{\sun}$. 
Alternatively, the maser emission could result from the
amplification of a radio jet by foreground circumnuclear molecular
gas. In this case, the high blueshift of the maser emission might mean
that the masing region is moving outward away from the molecular gas
surrounding an active nucleus.
\end{abstract}
%



\section{INTRODUCTION}
VLBI observations of water megamasers on (sub-)pc scales in active
galaxies have proven to be a basic tool for investigating the central
regions of active galactic nuclei (AGN). The most impressive result
from this kind of research was the discovery of a thin, nearly edge-on
rotating disk around a super massive black hole at the center of
NGC\,4258 \cite{miyo95, linc95}.  The VLBI observations revealed the
size and the shape of the slightly-warped maser disk with a central
radio jet outflowing perpendicularly to the disk plane
\cite{herr98}. VLBI imaging of the water megamasers in NGC\,1068,
NGC\,4945 and the Circinus galaxy showed the possible existence of
super massive objects at their nuclei similar to NGC\,4258
\cite{linc96, linc96b, linc97, linc01}. According to the most recent
result, in the Seyfert 2 galaxy IC\,2560, the barely resolved maser
spots which constitute the systemic emission show a linear velocity
gradient along the north-south elongation and are accompanied by
continuum emission, probably suggesting a super massive black hole
inside a rotating disk \cite{naka98, ishi00}.
\indent In the Seyfert 2 galaxy NGC\,1068, in addition to the nuclear
disk-maser, non-disk maser emission was found some 30 pc away along
the jet from the radio nucleus, that might be the result of
interaction between the jet and the interstellar gas \cite{gall96}. In
the elliptical galaxy NGC\,1052, the interpretation of non-disk masers
was introduced to explain the water masers seen lying along the jet
\cite{clau98}. 
\indent Water megamasers seem to fall into two categories: nuclear
masers residing in a (sub-)parsec scale disk around an active nucleus
and non-nuclear masers which are seen significantly further from the
nucleus. The latter can also be classified into two types, that is,
jet masers associated with radio jets and outflow masers -- a wind
component discovered in Circinus \cite{linc01}. The study of non-disk
maser emission is equally important for the investigation of the
overall structure surrounding an active nucleus. In order to
understand the megamaser phenomenon, and to deduce the general
characteristics of the central parsecs of AGN, such as black hole
masses and accretion rates, it is important to investigate as many
megamasers as possible with VLBI imaging at milliarcsecond angular
resolutions. So far, more than 20 galaxies are reported to contain
megamasers (e.g., Braatz et al. 1997, Falcke et al. 2000). Most of
them, however, show weak intensities, and some are thus not observable
with VLBI \cite{mora99}. \\
\indent NGC\,5793 is an edge-on disk galaxy, at a distance of 46 Mpc,
with a compact nucleus seen in radio continuum emission
\cite{gard92}. Baan et al.  (1998) classified the nucleus as being a
Seyfert 2 from its optical emission.  Hagiwara et al. (1997), using
the Nobeyama 45-m telescope, first detected the water maser emission
with highly Doppler-shifted satellite features, displaced by 245 \kms
on either side of the systemic velocity ($V_{\rm sys, \, LSR}$ = 3442
\kms; Palumbo et al. 1983) of the galaxy.  The total isotropic
luminosity of the maser was estimated to be 106 \lsun .  Fig.~1 shows
a new spectrum of blueshifted maser features measured with the Parkes
64-m telescope. There may be a suggestion of features near the
systemic velocity, which needs confirming. The highly redshifted
features at \vlsr = 3677 \kms, which were initially detected with the
45-m remain undetected since February 1996.  From 1996 we began VLBI
observations using the NRAO\footnote{~The National Radio Astronomy
Observatory is a facility of the National Science Foundation operated
under cooperative agreement by Associated Universities, Inc.} VLBA and
VLA, investigating the spatial distribution of the water maser
emission and the relationship between the water megamasers and the
nuclear radio activity of NGC\,5793. \\ 
\indent In this paper we describe the VLBA observations in Section 2,
present new images of the water vapour maser line and continuum
emission in the central parsecs of NGC\,5793 in Section 3, and discuss
the nature of the water maser in Section 4. We adopt a distance of 46
Mpc to NGC\,5793, corresponding to a scale of 1 milliarcsecond~(mas)
$\approx$ 0.23 pc.
\section{VLBA OBSERVATIONS AND DATA ANALYSIS}
\subsection{\it 22 GHz Observations}
\begin{figure}
\plotone{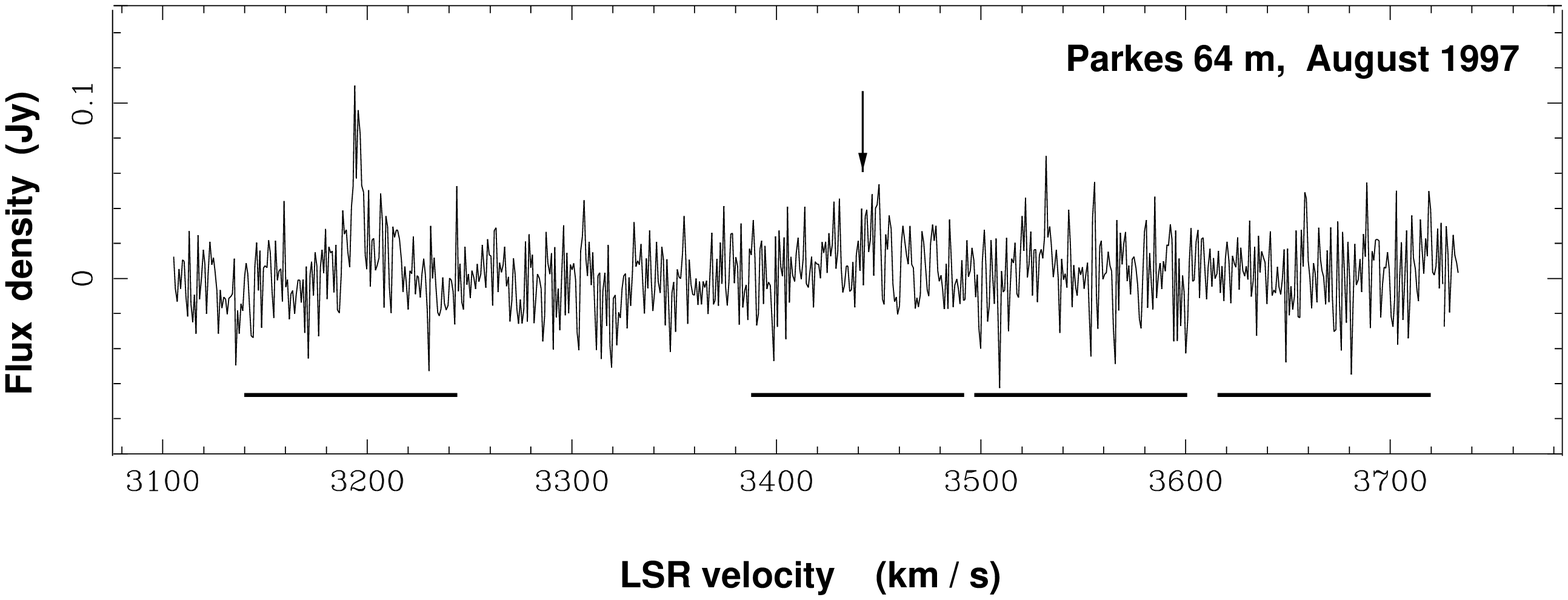} \figcaption[f1.eps]{Spectrum of the \hho maser
towards the center of NGC\,5793, observed with the Parkes radio
telescope of the CSIRO. The spectrum is averaged over three days from
18-20 August 1997. Accuracy of the flux density is about 10 \%. A
downward arrow indicates the adopted systemic velocity of the galaxy,
\vlsr = 3442 \kms. Solid bars indicate the velocity ranges covered by
four IFs which we observed with the VLBA in May 1998. \label{fig1}}
\end{figure}

We observed the 6$_{16} - 5_{23}$ maser line at 22 GHz (rest
frequency: 22.23508 GHz) toward NGC\,5793 for 8 hours on 1998 May 10
using the VLBA and the phased VLA. The use of the VLA as the eleventh
array element was critical to the observation because the peak flux
density of the maser emission is known to be very weak ($\sim$ 0.1
Jy). The data were recorded in left circular polarization with 2~bit
sampling in four 8 MHz IF bands of 128 spectral channels each
providing a velocity coverage of 108 \kms.  The four IF bands were
centered on \vlsr = 3190, 3442, 3551, and 3667 \kms. The observed
velocity ranges are displayed in a single-dish spectrum in
Fig.~\ref{fig1}.  The channel spacing in each band was 0.83 \kms
before averaging. We observed in a phase-referencing mode, using a
bright calibration source 1507$-$168 about 2 degrees away from
NGC\,5793. The total cycle time was two minutes; 1-minute scans on
1507$-$168 and NGC\,5793 alternated. The total amount of time spent
integrating on the target source was about 150 minutes. 4C\,39.25 and
3C\,345 were used for both bandpass and amplitude calibration. All the
data were correlated in NRAO Socorro and calibrated and reduced with
AIPS. The fringe-fit solutions (phase, rate, and delay) obtained from
the calibrator were applied to the IFs containing both the line plus
continuum emission. Five spectral channels were averaged together to
4.2 \kms to obtain reasonable sensitivities.  The positions of the
maser emission on the sky were determined by 2-D Gaussian fitting to
the maps. To make a continuum map, we averaged all spectral channels
in a line-emission-free IF, centered on 3667 \kms. \\ 
\indent The continuum and line maps were both made from differently
weighted data to get reasonable sensitivities, resulting in a
synthesized beam size of 2.8 $\times$ 1.8 mas for the continuum map,
and 1.8 $\times$ 1.0 mas for the line map.  The rms noise levels are
1.3 mJy \bm\ for the continuum map produced from a single bandwidth of
8~MHz and 4.3 mJy \bm\ for the continuum-subtracted line map for each
averaged channel. The continuum emission was subtracted only in the
region close to C1(C).
\subsection{\it Observations  at 1.7, 8.4, and 15 GHz}
The 1.7, 8.4, and 15~GHz continuum observations were made on 1999
October 10, using 8 elements of the VLBA. The VLBA stations
Brewster~(BR) and Saint Croix~(SC) did not take part in the
observations. The four 8 MHz IF bands with left circular polarization
were recorded using 2 bit sampling across a total bandwidth of 32
MHz. We used the frequency agility of the VLBA to switch rapidly among
the three frequencies to obtain better ($u$, $v$) coverages with
optimal observing time. The observing frequencies were switched
typically every 7 --10 minutes between 8.4 and 15~GHz.  The 1.7 GHz
observation was inserted among the other two frequencies about every
40 minutes. The 8.4 and 15~GHz observations were made using the
phase-referencing method, with the same calibrator as at 22~GHz. At
8.4 GHz a total cycle time of about 4 minutes was used for the
calibrator and the galaxy; 3-minute scans on the calibrator and
1-minute scans for the target source were employed. At 15~GHz,
1-minute scans on the calibrator and the target source were
alternated. The total integration time spent on the target source at
each frequency ranged from 60 to 120 minutes. The correlation of all
the data was performed at Socorro. The data were calibrated using
standard AIPS software.  The amplitude calibration was made with
observations of 4C\,39.25 and 3C\,345. The centroid position of the
8.4 GHz continuum source was determined from the phase-referenced
map. Finally, self-calibration was used on the data at 1.7 and 8.4 GHz
to improve the sensitivity of images with DIFMAP.  \\ 
\indent The maps were made from uniformly weighted data and the
resultant synthesized beam sizes are 23 $\times$ 3.7 mas at 1.7 GHz,
7.0 $\times$ 2.7 mas at 8.4 GHz, and 2.9 $\times$ 0.6 mas at 15~GHz.
The rms noise levels are 0.86 mJy \bm\ for the 1.7 GHz map, 0.35 mJy
\bm\ for the 8.4 GHz map, and 0.82 mJy \bm\ for the 15~GHz map at
total bandwidths of 32~MHz.
\subsection{\it Analysis of Positional Errors}
The positional errors ($\Delta$\,$\theta$) in the VLBI images, dominated by statistical noise, were estimated from the synthesized beam size ($\theta _{b}$) divided by the signal-to-noise ratios (SNR). When we estimate relative positional errors between line and continuum maps, they are approximately given by the the equation
\begin{displaymath}
\Delta \,\theta_{mc} \simeq \sqrt{ \left( \frac{\theta _{bm}}{2SNR_{m}} \right)^{2} + \left( \frac{\theta _{bc}}{2SNR_{c}} \right)^{2}}   ,
\end{displaymath}
where $m$ and $c$ represent maser and continuum emission images. Thus, the relative positional error between one of the maser lines at the \vlsr = 3194 \kms feature and the 22 GHz continuum peak C1(C) is $\theta _{mc}$ $\sim$ 0.22 mas.\\
\indent
The common reference positions (0, 0) (continuum peak C1(C)) were determined at various frequencies by phase-referencing to the compact source, 1507$-$168. This same calibrator was used for all the observing epochs and there seemed to be neither structural nor positional changes between the two epochs, May 1998 and October 1999. The alignment of the reference positions between the 22 and 8.4 GHz continuum maps was determined from the phase-referenced maps. Assuming that the position error of the common reference source and any structural changes of the reference source between epochs are negligible, the relative positional error between 22 GHz ($k$) and 8.4 GHz ($x$) is estimated as the following,
\begin{displaymath}
\Delta \,\theta_{mcx} \simeq \sqrt{ \left( \frac{\theta _{bm}}{2SNR_{m}} \right)^{2} + \left( \frac{\theta _{bcx}}{2SNR_{cx}} \right)^{2} + \left( \frac{\theta _{brk}}{2SNR_{rk}} \right)^{2} + \left( \frac{\theta _{brx}}{2SNR_{rx}} \right)^{2}  }   ,
\end{displaymath}
where $c$ and $r$ represent continuum images of a target source  and a phase-reference source. In our analysis, the third and fourth term in the formula are negligible as SNR of reference calibrator images are better by on the order of 10 -- 100~than those of masers or 22 and 8.4 GHz continuum images of NGC\,5793. Consequently, the relative uncertainty of the maser position with respect to the reference point at 8.4 GHz in Fig.~\ref{fig3} is $\sqrt{ \left(\theta _{bm} / 2SNR_{m} \right)^{2} + \left(\theta _{bcx} / 2SNR_{cx} \right)^{2} }$ = $\sqrt{ \left( 2/ (2 \cdot 11) \right)^{2} + \left( 7/ (2 \cdot 22) \right)^{2} }$ $\sim$ 0.2 mas. (The inserted values were adopted from the phase-referenced maps without self-calibration.) 
\section{RESULTS}
\subsection{\it Continuum  Emission}
Fig.~2 and Fig.~3 show radio continuum images at 22 and 8.4 GHz, from
which we can identify at least four continuum components C1(C), C2(E),
C2(C) and C2(W). The naming of the components follows the convention
used in Hagiwara et al. (2000). The 22 GHz peak is coincident with
C1(C) in the 8.4 GHz map to an accuracy of 0.25 mas, as estimated in
the previous section. The basic morphological structure is quite
similar to those at 1.7 and 5.0 GHz given in Fig.~2 and Fig.~3 of the
paper by Hagiwara et al. (2000), but at both 15 and 22 GHz only C1(C)
is detected.  Table~\ref{tbl-1} lists the Gaussian-fitted peak flux
densities of the major compact continuum sources identified at the
five frequencies, and Table~\ref{tbl-2} lists the spectral indices of
each component.  The flux density and spectral indices were obtained
by convolving with the same beam size corresponding to that of the 1.7
GHz map in 1996 to avoid resolution effects. The poor ($u$, $v$)
coverage for the 1.7 GHz observing session in Oct. 1999 resulted in
relatively inaccurate component flux densities in comparison with
those in Nov. 1996. Fig.~\ref{fig4} shows spectra from four major
components imaged in Fig~\ref{fig3}. C1(C) shows steep spectral
indices $\alpha < - 0.70$ (using the convention that S$_{\nu}$
$\propto$ $\nu^{\alpha}$, where S$_{\nu}$ is the flux density at
frequency $\nu$) from 1.7 GHz to 15~GHz but a flat index $\alpha \,
\simeq\, -0.02$ between 15~GHz and 22~GHz. C2(C) and C2(W) have
similar spectra: $\alpha = -0.88$ and $-0.99$ between 5.0 and 8.4 GHz,
and $\alpha = -0.46$ and $-0.72$ between 1.7 and 5.0 GHz. The compact
component C2(E) that lies within the jet extension and shows an
inverted spectrum $\alpha \simeq +0.13$ between 1.7 GHz and 5.0
GHz. The continuum peak of C2(E) was not clearly seen at 1.7 GHz in
both epochs, but was seen in a 1.4 GHz map obtained in Dec. 1996
\cite{pihl00}.

\subsection{\it Water Maser Emission}
\begin{figure}
\plotone{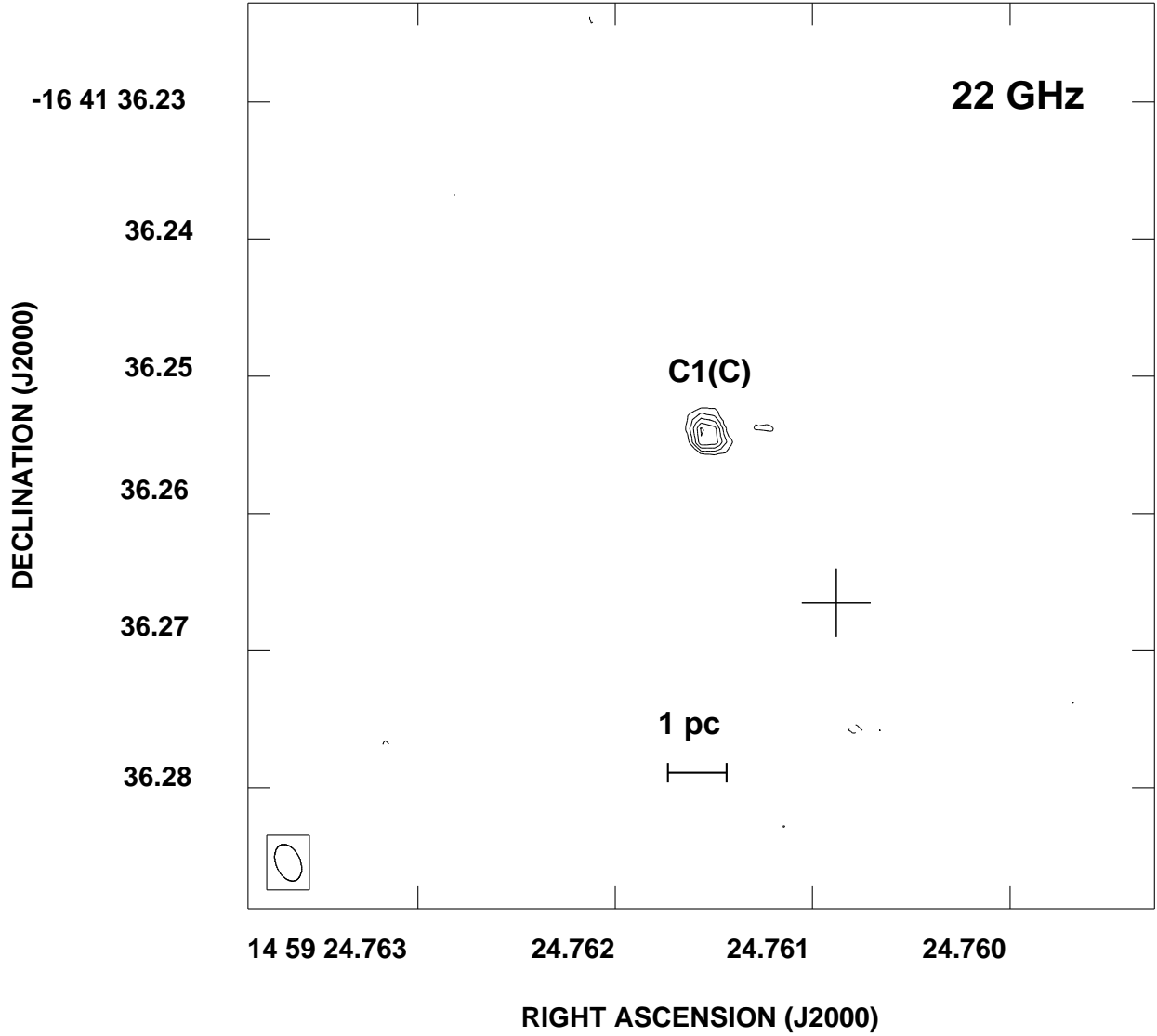} \figcaption[f2.eps]{Radio continuum image at 22 GHz
with the VLBA. The levels of contours are -3, 3, 4, 5, 6, 7$\sigma$ (1
$\sigma$ = 1.3 mJy \bm). The peak flux density is 9.2 mJy \bm. The
synthesized beam of 2.8 $\times$ 1.8 mas at P.A. = 23$\,^{\circ}$ is
plotted in the left-hand corner. The cross marks the Gaussian-fitted
center of a maser feature at \vlsr = 3194 \kms. The relative alignment of the
maser and continuum is accurate to 0.22 mas.
\label{fig2}}
\end{figure}

\begin{figure}
\plotone{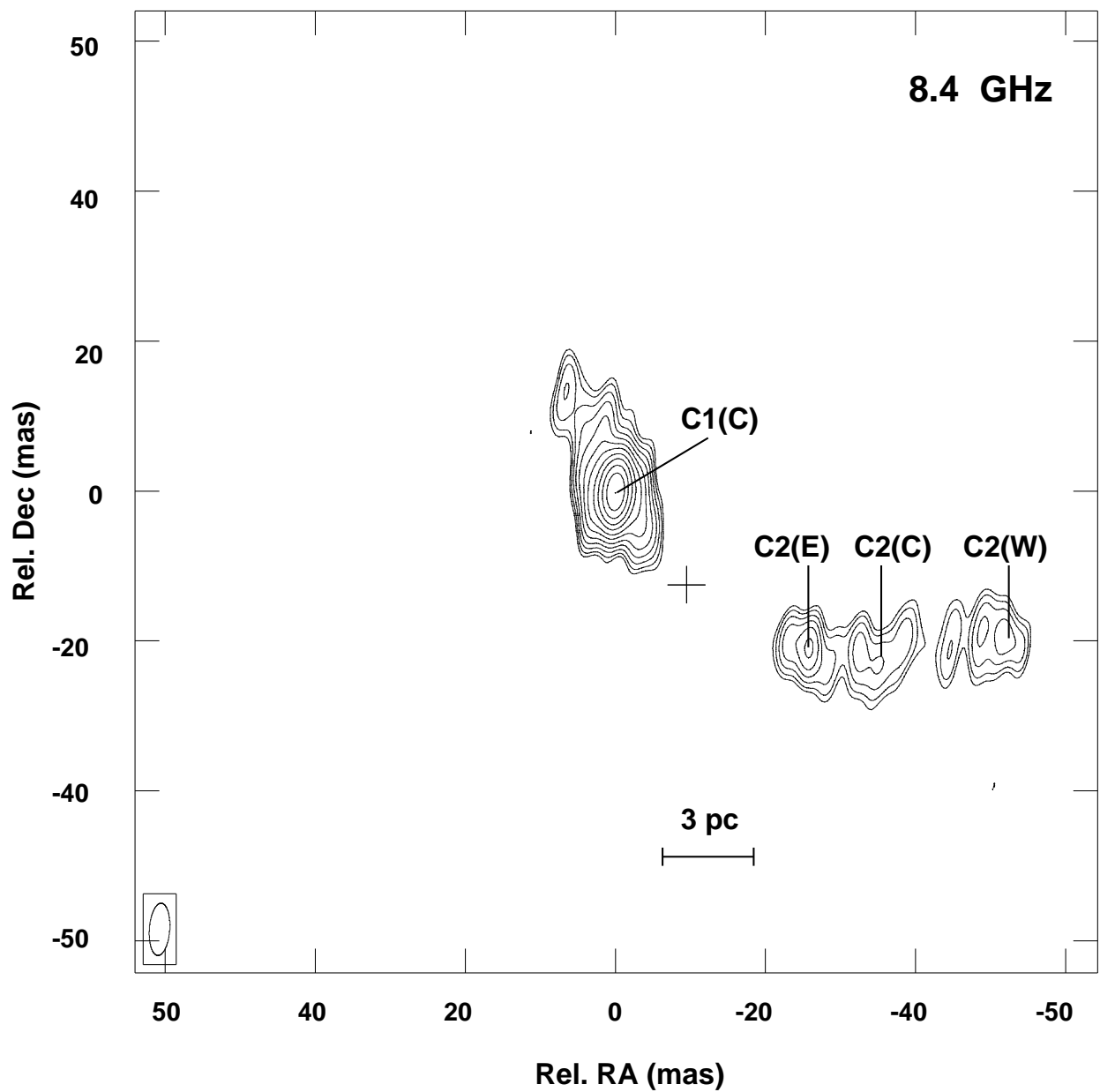} \figcaption[f3.eps]{8.4 GHz radio continuum image of
the nuclear region made with the VLBA. The contour levels are $-2.7,
2.7, 3.7, 5.2, 7.2, 10, 14, 19, 27, 37, 52, 72, ~{\rm and}~100\,\%$ of
the peak flux density of 51.1 mJy \bm. The synthesized beam of 7.0
$\times$ 2.7 mas at P.A. = -- 4.7$\,^{\circ}$ is shown in the
left-hand corner. The position of the maser feature at \vlsr = 3194
\kms is marked by a cross. The center position (0, 0) in the map
coincides with that of the 22 GHz continuum peak in Fig.~2. The
relative accuracy of the 8.4 GHz continuum and 22 GHz maser positions
is $\sim$ 0.2 mas.\label{fig3}}
\end{figure}

\begin{figure}
\plotone{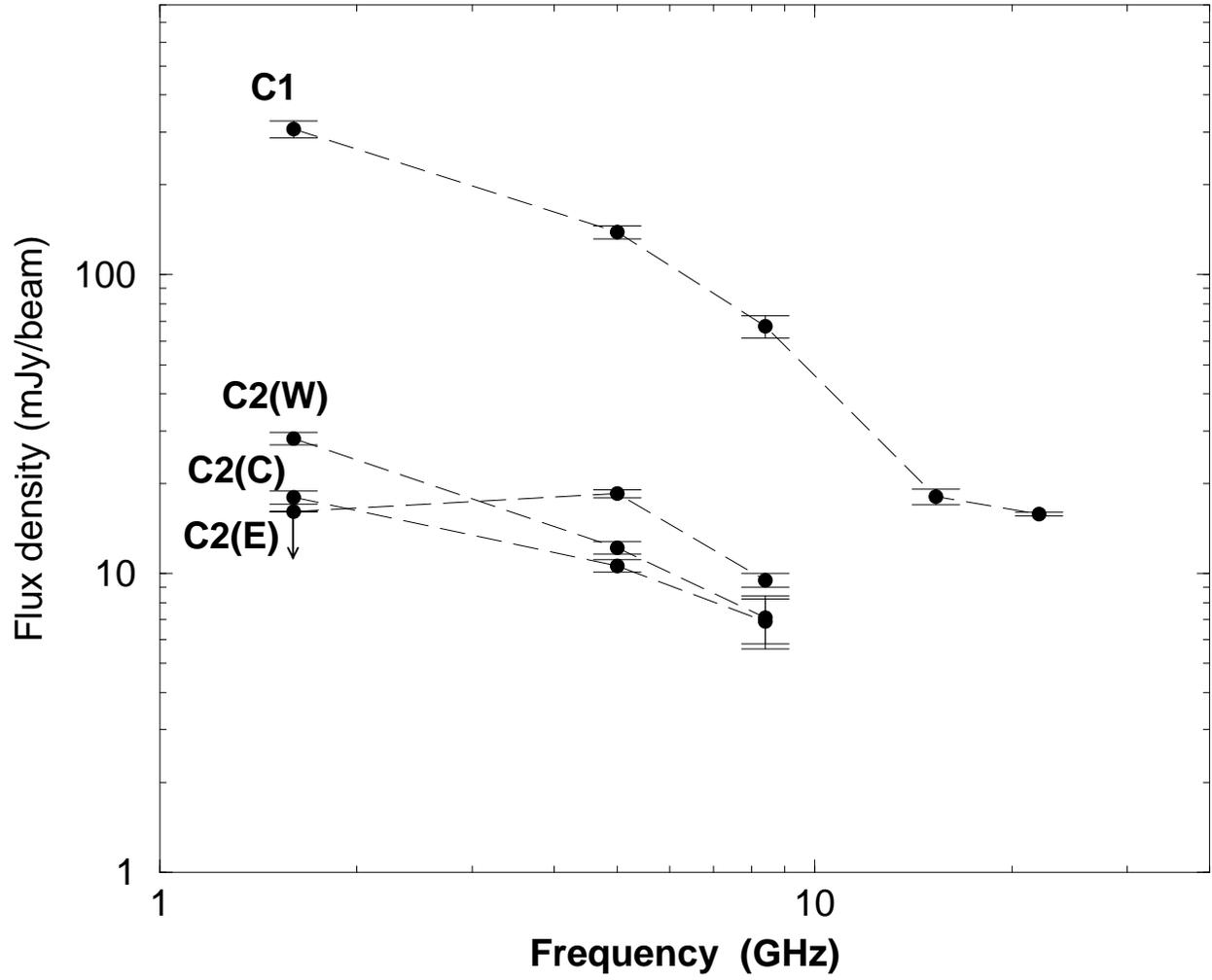} \figcaption[f4.eps]{Continuum spectra of the four
compact components C1(C), C2(E), C2(C), and C2(W) at 1.7, 5.0, 8.4, 15
and 22 GHz. The flux density of C2(E) at 1.7 GHz is an upper
limit. \label{fig4}}
\end{figure}

\begin{figure}
\plotone{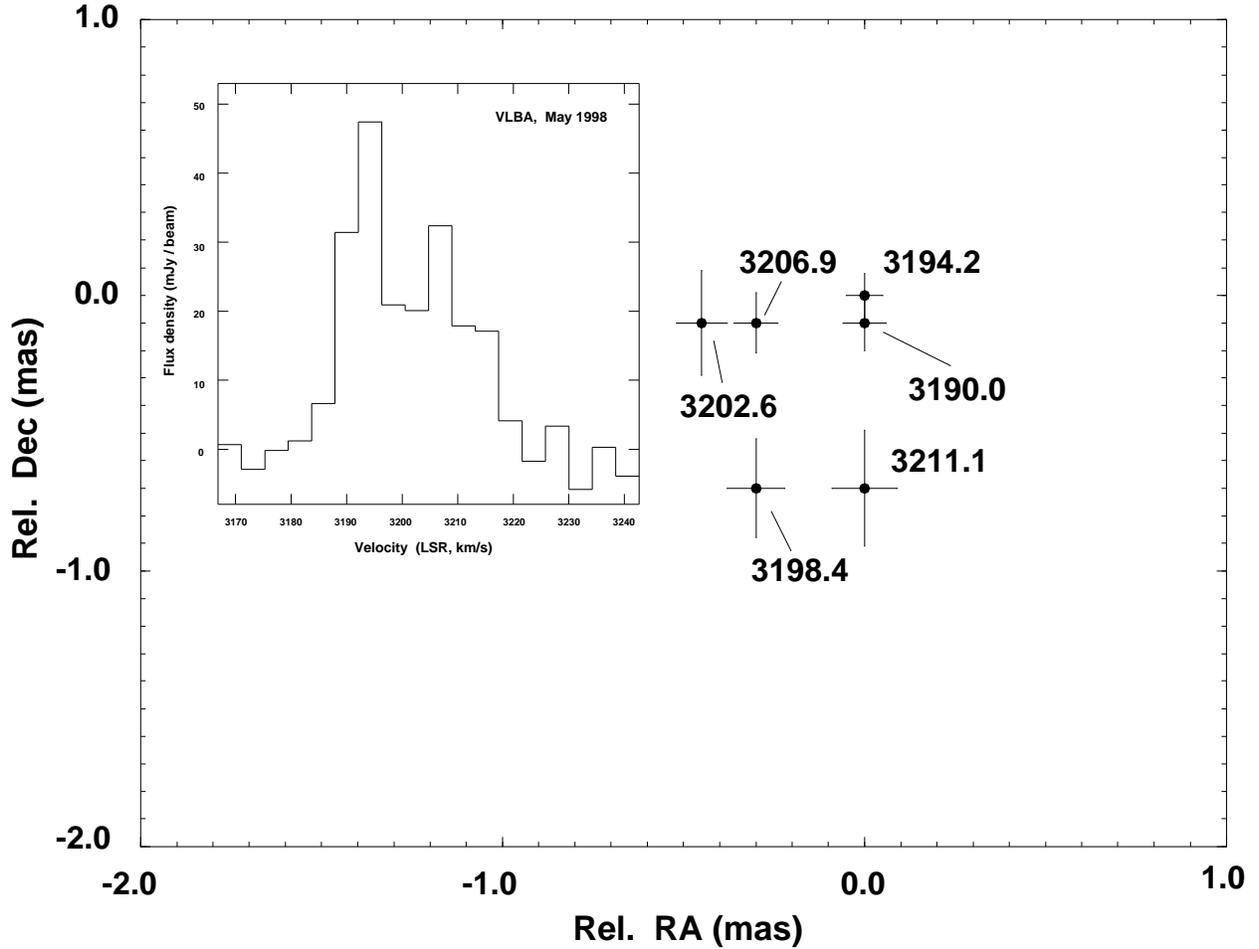} \figcaption[f5.eps]{Distributions of water maser
emission. The position of each component was estimated by 2-D
Gaussian-model fitting, and error bars indicate total positional
uncertainties of 1$\sigma$. The point (0, 0) is referred to position
of the crosses in Fig. 2 and Fig. 3. The maser spots are labelled by
their velocities in \kms. The inset shows the VLBA spectrum of the
blueshifted maser emission. Five spectral channels were averaged
to 4.2 \kms (channel spacing before averaging is 0.84 \kms). Accuracy
of the flux density scale is typically 5 \%. \label{fig5}}
\end{figure}

We detected maser features that are blueshifted with respect to the
systemic velocity of the galaxy with the VLBA.  The maser features are
seen in the averaged channels ranging from \vlsr = 3190 \kms -- 3210
\kms.  Inset into Fig.~\ref{fig5} is a blueshifted spectrum that shows
the total flux density obtained from our VLBA observations. The 
spectrum appears to be a scaled-down version of the the single-dish 
spectrum, which suggests that about 50 \% of the peak flux density is
detected in VLBA observations. This might be explained by the
intrinsic intensity variability of the water emission or by a loss of
coherence during phase referencing \cite{cari99} or, of course, by a
combination of both. Given the total maser intensity (0.78 Jy \kms)
estimated from the VLBA spectrum shown in the inset in
Fig.~\ref{fig5}, the isotropic luminosity of the blueshifted emission
is 38 \lsun, which is comparable to that of the redshifted emission
(22 \lsun) in NGC\,4258 \cite{naka95}.  Fig.~\ref{fig5} shows the
spatial distribution of the water maser spots, where the position of
each maser feature was determined by Gaussian fitting for each
emission peak in the deconvolved images that was detected above
4$\sigma$ level.  Our analysis showed that the detected maser emission
is split into two clusters with a separation of $\sim$ 0.7 mas, or 0.16
pc, but there is no discernible velocity structure. Each maser clump,
assuming a nominal size of 0.2 mas, yields a lower limit to the
brightness temperature, T$_{\rm b}$ $\sim$ 10$^9$ K. The centroid
position of the brightest maser feature centered at \vlsr = 3194 \kms\
is marked with a cross in Fig.~\ref{fig2} and Fig.~\ref{fig3} in which
the continuum maps are superposed. No continuum emission was detected
at the maser position at any of our observed frequencies. The maser
spot is located 15.5 mas or 3.6 pc from the unresolved radio continuum
peak C1(C) at 22 GHz, which is coincident with a reference point (0, 0) in the 8.4 GHz map in Fig.~\ref{fig3}. From VLA-A observations with
a beam size of 130 $\times$ 80 mas (30 $\times$ 18 pc) in 1997
January, this blueshifted maser emission remained unresolved, and was
coincident with the unresolved continuum peak C1(C), supporting our
VLBA results in this paper. The peak of C2(E) at 8.4 GHz is about 18
mas, some 4 pc away from the maser spot, suggesting that the maser
clusters are located nearly at the midpoint of the line joining C1(C)
and C2(E). \\
\indent We found no velocity features of the water maser corresponding
to the velocities of the HI and OH absorptions, observed with the VLA
by Gardner et al. (1986).
%
%
%
%
%
%
%
\section{DISCUSSION}
\subsection{\it Location of the Nucleus}
One might ask where the nucleus of NGC\,5793 lies? In order to
understand the kinematics of the nuclear region in the galaxy we need
to address this question. Component C1(C) shows a flat-spectrum at
higher frequencies, at which it is not resolved, suggesting that it
might be the 'true' nucleus. On the other hand, the `true' nucleus may
lie in the continuum emission gap between C1(C) and C2(E) where the
maser spots are located.  This would mean that the radio continuum
nucleus could be highly obscured by a screen of dense ionized gas. In
some megamasers like NGC\,1068 \cite{linc96}, the Circinus galaxy
\cite{linc01}, and possibly NGC\,4945 \cite{linc97}, the existence of
maser disks has been confirmed but continuum emission has not been
detected in the vicinity of the water masers. \\
\indent The measured surface brightness of C2(E) is at least 3.2
$\times$ 10$^8$ K and could in principle be as high as the $\sim$
10$^{10}$ K required for synchrotron self-absorption \cite{kell69}. It
is conceivable that component C2(E) was produced by a strong shock
with the circumnuclear gas. A jet ejected from the nucleus could be
bent and deflected near C2(E) and extend further out to component
C2(W).  Alternatively, component C2(E) could be free-free absorbed by
dense ionized foreground gas. The spectral index of +0.13 between 1.7
and 5.0 GHz is not steep enough for free-free absorption, but as it is
a lower limit, it is not inconsistent with such an hypothesis. If
C2(E) is a free-free absorbed component, it could be the inner part of
the disk/torus. However, at this stage we $conservatively$ consider
that the spectral turnover of C2(E) is due to synchrotron
self-absorption caused by the collision of a jet from the central
engine with a dense gas in the circumnuclear region. \\ 
\indent There might be a change in flux density during the two
observing sessions between Oct. 1999 at 1.7 GHz and Oct. 1997 at 5.0
GHz. If so, the determination of the spectral index of C2(E) is
unreliable. However, the flux density of C2(E) for 1.7 GHz in
Nov. 1996 is quite similar with that in Oct. 1999, hence it is less
likely that there was significant flux density variation from 1996 to
1999 as for C2(E). In any case, there is no straightforward
interpretation for the location of the 'true' core. For this reason it
is impossible to decide at present whether the water emission in
NGC\,5793 is a nuclear or non-nuclear maser.
\subsection{\it Origins of the Water Maser}
\begin{figure}
\plotone{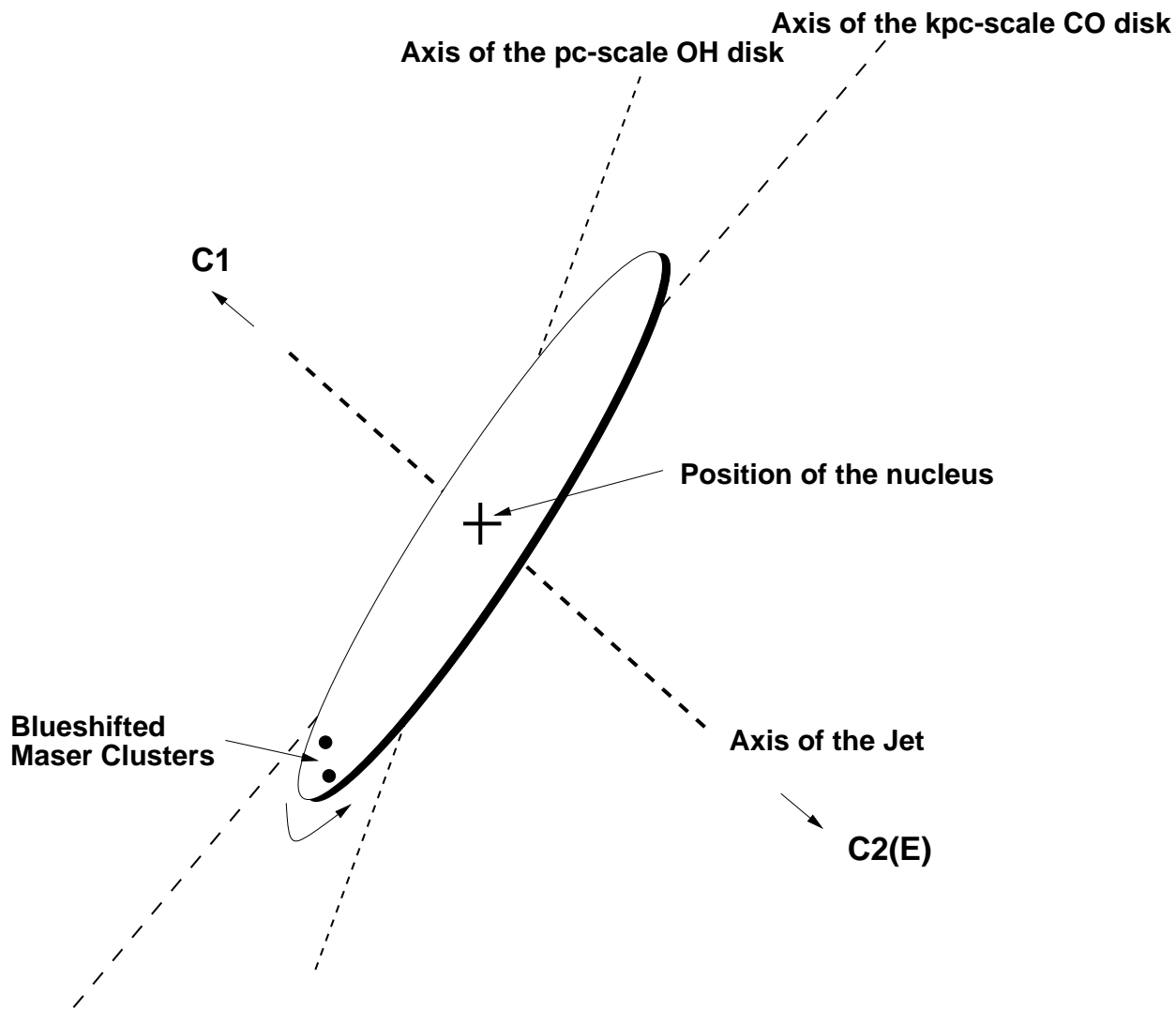} \figcaption[f6.eps]{Schematic view of the nuclear
region of NGC 5793.  The position of the nucleus is assumed to lie at
the midpoint on the line joining C1 and C2(E). The blueshifted maser
clusters are located on the approaching side of a postulated edge-on
rotating disk. The P.A. of the disk lying nearly perpendicular to the
jet is well aligned with the kilo-parsec-scale disk observed in CO
emission, and the pc-scale disk imaged in OH absorption
\cite{hagi00}. An arrow denotes the sense of the rotation of the maser
disk. The sense is the same as that of the OH disk, but reversed when
compared to that of the CO disk.
\label{fig6}}
\end{figure}

Interpretation of the geometry of the detected maser emission in
NGC\,5793 is not straightforward, because the blueshifted features are
located between continuum peaks and the velocity fields do not show
any systematic trends. However, we discuss two possibilities to explain
the nature of the maser in the galaxy:
\begin{itemize}
\item[1] The water maser in NGC\,5793 results from the amplification
of a radio jet by the circumnuclear molecular gas foreground to the
jet.
\item[2] The maser lies in a in a compact molecular disk rotating
around the galactic nucleus, like that in NGC\,4258.
\end{itemize}
\indent Support for the validity of the first model comes from the
presence of molecular (OH) gas with a parsec-scale distribution seen
in absorption against the continuum structure with the VLBA
\cite{hagi00}. The brightness of a putative radio jet component may
not be sufficient to be detected by VLBI in NGC\,5793 at 22~GHz. The
brightness temperature of any undetected 22 GHz continuum source near
the maser spots is less than $\sim$ $10^7$ K, assuming that the peak
flux density is 6.5 mJy \bm\ (5$\sigma$ noise level). The beam
averaged brightness temperature of the unresolved maser spot is 1.3
$\times$ 10$^8$ K, so that the maser gain would be a factor of $>$
10. Usually, this situation is seen in OH megamasers functioning as
low-gain amplifiers against background continuum sources. Non-nuclear
and non-disk masers have been seen in NGC\,1052 \cite{clau98} and
NGC\,1068 \cite{gall96}. The non-nuclear water maser known to exist in
NGC\,1068, located at 30 pc from the nucleaus along the jet, has not
been detected with VLBI, suggesting that the maser spot is resolved or
not intense enough for VLBI detection. In NGC\,1052 two maser spots
are seen lying 0.07 pc from the presumed core along a radio jet; they
show no evidence for disk structure \cite{clau98}. Neither of the
above cases seems to explain the maser in NGC\,5793 because it
exhibits different characteristics. If the water maser of NGC\,5793 is
emitted from the nuclear region, the large blueshift ($\sim$ 245 \kms)
of the maser emission implies that it traces molecular outflows moving
at high speed relative to the molecular disk surrounding the
nucleus. A model of a non-disk maser as described here also has
problems in explaining the single-dish spectrum of the emission.\\
\indent We now consider the the second possibility, namely that the
masers lie in a molecular disk rotating around the nucleus of the
galaxy as proposed by Hagiwara et al. (1997). Unfortunately, during
the period of observations the systemic and redshifted features became
weaker, and we could not measure the distribution and the velocity
structure of the water masers. If we assume the presence of rotating
disk, the blueshifted features could arise from the edge of the
disk. The emission has been detectable and does not show any velocity
drifts since its discovery in January 1996. In addition, the
peak flux density is less variable than other features. Such properties
are  similar to the characteristics of the blue- and
redshifted masers in NGC\,4258 \cite{linc95b}. By contrast, the peak
flux densities of features around the systemic velocity have been
particularly variable. Greenhill et al. (1995a) suggested the
possibility that the high-velocity features in NGC\,4258 are visible
for substantially longer times than the systemic ones because the
emission might be self-amplified in the long gain paths along the edge
of the disk. Considering these facts, the high-velocity blueshifted
features may trace only a part of the disk, possibly near the tangent
points. 
\indent With such a supposition in mind, we propose a model
(Fig.~\ref{fig6}) for the nuclear region of NGC\,5793. The location of
the nucleus is presumed to lie on the midpoint of the axis connecting
C1(C) to C2(E). This axis may mark the location of a two-sided jet.
The blueshifted maser features are offset from this axis by $\sim$ 4.5
mas, or 1.0 pc. The southern side of the rotating disk, where the
maser clumps lie, is approaching, while the northern side is
receding. The disk orientation is nearly perpendicular to the jet. In
this model, the inferred free-free absorbed source C2(E) lies within
the inner radius of the disk. We estimate that the total mass enclosed
within a radius of 1 pc is $\sim$ 1 $\times$ 10$^7$ $M_{\sun}$,
assuming a rotation velocity of 240 \kms and a disk inclination of
73$^{\circ}$ \cite{roth94}. The mass density is $\sim 2 \times 10^6 \,
M_{\sun} \, {\rm pc}^{-3} $, a value that is the smallest among the
megamasers observed with VLBI \cite{mora99}. The enclosed mass of
$\sim$ 10$^7$ $M_{\sun}$ is lower approximately by a factor of 10 than
that estimated from the OH absorption in the central region of $\sim$
10 pc \cite{hagi00}, because we adopted a smaller radius. The sense of
the disk rotation is opposite to that of the galactic disk, which has
a radius of $\sim$ 1 kpc and is observed in CO(J = 1 -- 0) emission,
but is the same as that of the OH absorbing gas disk, implying the
existence of an independent kinematical system within the central few
parsecs of the circumnuclear region. \\
\indent From recent single-dish measurements made in early 2000,
systemic features lying at \vlsr = 3430 \kms to 3520 \kms were found
to be flaring and variable in flux density (Henkel, private
communication), suggesting that the features might arise against the
continuum source at the nucleus along the line of sight, and that the
masing cloud is amplifying the background continuum source.
\section{SUMMARY}
We have observed the water megamasers and continuum emission in the
nuclear region of the Seyfert 2 galaxy NGC\,5793 with the
VLBA. Because of the faintness of the maser intensities, only the
blueshifted emission could be imaged.  This emission, which is located
3.6 pc southwest of the radio continuum source, is observed to consist
of two clusters with a separation of 0.7 mas (0.16 pc). No compact
continuum emission was detected near to the maser spot. The detected
water maser could be a part of a maser disk as in the case of
NGC\,4258. We propose a model in which the masers lie on the southern
edge (approaching side) of a rotating molecular gas disk whose
orientation is nearly perpendicular to the jet. Assuming the location
of the nucleus at a midpoint between C1 and C2(E) one can estimate a
binding mass on the order of 10$^7$ $M_{\sun}$. Alternatively, the
masers could result from the amplification of a radio jet by
foreground circumnuclear gas. The highly blueshifted velocity of the
maser features might mean that the masing cloud traces outflowing 
gas surrounding an active nucleus. \\ 
\indent The compact continuum source, C2(E) appears to have an
inverted spectrum between 1.7 and 5.0 GHz, assuming no intrinsic
variability, and the spectral turnover might be due to synchrotron
self-absorption caused by a shock in the jet owing to collision with
the dense gas, or a free-free absorbing screen in a part of an ionized
torus/disk. \\ 
\indent Further VLBI observations of the water masers and continuum
emission are crucial to clarify the dynamical structures of the maser
emission in relation to the active nucleus. The systemic and
redshifted velocity features flared in early 2000 and additional VLBA
observations, following up our present result, are already underway.
\acknowledgments
We are grateful to Drs. R. Porcas, E. Ros, C. Henkel, and A. Roy for
their helpful comments and suggestions. We also appreciate J. Conway
and Y. Pihlstr{\"o}m for providing us the 1.4 GHz image. YH
appreciates the efforts of P. Hall, E. Troup, and other staff members
at Parkes for their assistance in the observations and data
analysis. Part of the VLBA observations appearing in this paper were conducted
while PJD was an NRAO staff member. This research has made use of NASA's Astrophysics Data System Abstract Service.

\clearpage
\clearpage
%

%
\begin{deluxetable}{crrrrrr}
\tabletypesize{\scriptsize}
\tablecolumns{7}
\tablewidth{0pc} 
\tablecaption{\sc Flux Densities of Major Components  \label{tbl-1}}

\tablehead
{
\colhead{}    &  \multicolumn{4}{c}{Component Flux Densities} \\
\colhead{}    &  \multicolumn{4}{c}{(mJy \bm)} \\
\colhead{Frequency}& \cline{1-4}
\colhead{Band} & \colhead{C1(C)}  & \colhead{C2(E)} & \colhead{C2(C)}  &\colhead{C2(
W)} & \colhead{Epoch} & \colhead{Reference} \\
\multicolumn{1}{c}{(GHz)} &\colhead{} 
}
\startdata
1.4\nodata &304&16.3&15.6&23.1& Dec. 96&\colhead{a}\\
1.7\nodata  &308&$<$16.1\tablenotemark{d}&18.0 & 28.3&Nov. 96&\colhead{b}  \\
        &274 &$<$ 15.2\tablenotemark{d}  &13.9  &14.7  &Oct. 99&\colhead{c}  \\
5.0\nodata  &139& 18.5&10.6&12.2& Oct. 97&\colhead{b}\\
8.4\nodata   & 64.5 &9.42 &6.73 &7.31 &Oct. 99&\colhead{c}  \\
15\nodata  & 16.4   &\nodata & \nodata& \nodata&Oct. 99&\colhead{c} \\
22\nodata  & 16.3  &\nodata&\nodata&\nodata &May 98&\colhead{c} \\
\enddata
\tablenotetext{a}{Y. Pihlstr{\"o}m, private communication}
\tablenotetext{b}{Hagiwara et al. 2000}
\tablenotetext{c}{This paper}
\tablenotetext{d}{An upper limit value}

\tablecomments{Flux uncertainties assume 5 persent calibration uncertainty added. All the flux densities are estimated with the synthesized beam at 1.7 GHz (12.4 $\times$ 4.4 mas).}
\end{deluxetable}
%
%
%
\begin{deluxetable}{crrrrrr}
\tabletypesize{\scriptsize}
\tablecolumns{5}
\tablewidth{0pc} 
\tablecaption{\sc Spectral Indices of Major Components  \label{tbl-2}}

\tablehead
{
\colhead{}    &  \multicolumn{4}{c}{Component Spectra, $\alpha$} \\
\colhead{}    &  \multicolumn{4}{c}{(S$_{\nu}$ $\propto$ $\nu^{\alpha}$)} \\
\colhead{Frequency Range}& \cline{1-4}
\colhead{(GHz)} & \colhead{C1(C)}  & \colhead{C2(E)} & \colhead{C2(C)}  &\colhead{C2(W)}
}
\startdata
\colhead{1.7\tablenotemark{a} $-$ 5.0}\nodata&$-$0.70&0.13&$-$0.46 & $-$0.72\\
\colhead{5.0 $-$ 8.4}\nodata&$-$1.5&$-$1.3&$-$0.88&$-$0.99\\
\colhead{8.4 $-$ 15}\nodata&$-$2.3&\nodata &\nodata&\nodata   \\
\colhead{15 $-$ 22}\nodata&$-$0.017& \nodata&\nodata&\nodata
\enddata

\tablenotetext{a}{Data from Nov. 1996}

\end{deluxetable}

\end{document}